\documentclass[aps,prl,showpacs,groupedaddress,superscriptaddress,twocolumn]{revtex4}

\usepackage{graphicx, amsmath, subfigure}
\usepackage{hyperref}
\usepackage{color}
\usepackage{float}
\usepackage{chngcntr}

\begin{document}

\preprint{APS/123-QED}


\title{Non-Abelian braiding of Dirac fermionic modes using topological corner states in higher-order topological insulator}

\author{Yijia Wu}
\affiliation{International Center for Quantum Materials, School of Physics, Peking University, Beijing 100871, China}

\author{Hua Jiang}
\affiliation{School of Physical Science and Technology, Soochow University, Suzhou 215006, China}

\author{Jie Liu}
\affiliation{Department of Applied Physics, School of Science, Xian Jiaotong University, Xian 710049, China}

\author{Haiwen Liu}
\affiliation{Center for Advanced Quantum Studies, Department of Physics, Beijing Normal University, Beijing 100875, China}

\author{X. C. Xie}
\thanks{Corresponding author: xcxie@pku.edu.cn}
\affiliation{International Center for Quantum Materials, School of Physics, Peking University, Beijing 100871, China}
\affiliation{Beijing Academy of Quantum Information Sciences, Beijing 100193, China}
\affiliation{CAS Center for Excellence in Topological Quantum Computation, University of Chinese Academy of Sciences, Beijing 100190, China}

\date{\today}

\begin{abstract}
We numerically demonstrate that the topological corner states residing in the corners of higher-order topological insulator possess non-Abelian braiding properties. Such topological corner states are Dirac fermionic modes other than Majorana zero-modes. We claim that Dirac fermionic modes protected by nontrivial topology also support non-Abelian braiding. An analytical description on such non-Abelian braiding is conducted based on the vortex-induced Dirac-type fermionic modes. The braiding operator for Dirac fermionic modes is also analytically derivated and compared with the Majorana zero-modes. Experimentally, such non-Abelian braiding operation on Dirac fermionic modes is proposed to be testified through topological electric circuit.
\end{abstract}

\pacs{05.30.Pr, 03.65.Vf, 03.75.Lm, 03.67.Lx}
\maketitle



\textit{Introduction.} Higher-order topological insulator \cite{HOTI_ScienceAdvances, 2D_SSH_model_science, HOTI_inversion_symmetry, HOTCI_PRX} (HOTI) has been drawing great attention for possessing novel boundary states including topological corner state \cite{corner_state_HuaJiang, 2D_SSH_model_science, HOTCI_Cn_symmetry} and topological hinge state \cite{HOTI_ScienceAdvances, HOTI_inversion_symmetry}. As the bound state localized at the defect or spatial boundary of the 1D gapped topological edge state, topological corner state can be viewed as an incarnation of the celebrated Jackiw-Rebbi zero-mode \cite{J-R_PRD, SSH_model, DanielLoss_J-R_in_QSH, YijiaWu} in 2D or 3D condensed matter systems. As one of the fascinating properties of Jackiw-Rebbi zero-mode, the charge fractionalization has also been widely investigated for topological corner state in HOTI \cite{HOTI_ScienceAdvances, 2D_SSH_model_science, HOTI_inversion_symmetry, HOTCI_Cn_symmetry, ChangYu_Hou_fractionalization_graphene}. Another fascinating property of Jackiw-Rebbi zero-mode is its non-Abelian statistics \cite{DanielLoss_J-R_non-Abelian-1, DanielLoss_J-R_in_QSH, YijiaWu} that a geometric phase highly related to the nontrivial topology is accumulated \cite{vonOppen_braiding, Pachos_review, Ezawa_Majorana_circuit_braiding_1} during the braiding. However, the non-Abelian braiding of the topological corner state in HOTI has only been investigated when superconducting pairing is presented \cite{Majorana_corner_braiding, Ezawa_Majorana_circuit_braiding_1, Ezawa_Majorana_circuit_braiding_2}, in which the topological corner states are self-conjugate and therefore are actually Majorana zero-modes (MZMs) \cite{Kitaev's_chain, YijiaWu, MatthewFisher_T-junction, IvanovPRL2001}. Nevertheless, in practice, 
such Majorana condition is absent for a number of platforms supporting topological corner states such as higher-dimensional Su-Schrieffer-Heeger (SSH) lattice \cite{SSH_model, 2D_SSH_model_science} and graphene-like structure \cite{ChangYu_Hou_fractionalization_graphene, corner_state_HuaJiang}. In view of this, it is of significant importance investigating the possible non-Abelian statistics of topological corner states in an HOTI without superconduting pairing term.

Though first raised in the field of condensed matter physics, the HOTI has only been confirmed in few condensed matter materials such as bismuth \cite{HOTI_bismuth}. In spite of this, the idea of HOTI has already been realized in other areas including phononic crystal \cite{HOTI_phononic_nature, HOTI_phononic_nature_physics, HOTI_phononic_nature_materials_1, HOTI_phononic_nature_materials_2}, photonic crystal \cite{HOTI_photonic_nature, HOTI_photonic_PRL_1, HOTI_photonic_PRL_2, corner_state_HuaJiang}, and topological electric circuit composed of inductors and capacitors \cite{2D_SSH_circuit_1, 2D_SSH_circuit_2, 3D_SSH_circuit_experiment, Ezawa_LC_circuit}. 
A great advantage of topological electric circuit is that through multiple connections of the circuit nodes, lattice in three-dimension or even higher-dimensions could also be constructed \cite{RuiYu_4DTI, Ezawa_LC_circuit, 3D_SSH_circuit_experiment}. Moreover, the tight-binding parameters in circuit lattice can be locally modulated through variable inductors or capacitors, making topological circuit an ideal scheme simulating time-dependent lattice Hamiltonian.

In this Letter, we first perform a numerical simulation demonstrating the non-Abelian braiding properties of HOTI's topological corner states based on a 2D SSH model. Due to the absence of the Majorana condition, such topological corner states are actually Dirac-type fermionic modes protected by the topology. 
Such localized Dirac fermionic modes in 2D topological system also appear as the vortex-bounded states in quantum anomalous Hall insulator (QAHI), in which the non-Abelian nature of the topologically protected Dirac fermionic modes is proved in an analytical way. The braiding operator for these Dirac fermionic modes is different from the MZM \cite{IvanovPRL2001} and the bosonic mode \cite{nonAbelian_light}, although their geometric phases accumulated during the braiding have the same form. Finally, an experimental proposal is raised to simulate the braiding of HOTI's topological corner states 
in a topological circuit with variable inductors.



\textit{Non-Abelian braiding of topological corner states.} As the minimal model of HOTI, the two-dimensional generalization \cite{2D_SSH_model_science} of the SSH model \cite{SSH_model} is an ideal platform demonstrating the non-Abelian braiding properties of HOTI's topological corner states. The Hamiltonian of the 2D SSH model $h_{\mathrm{SSH}} (\mathbf{p}) = (\gamma + \lambda \cos p_x) \tau_x \sigma_0 - \lambda \sin p_x \tau_y \sigma_z - (\gamma + \lambda \cos p_y) \tau_y \sigma_y - \lambda \sin p_y \tau_y \sigma_x$ \cite{2D_SSH_model_science} can be discretized in a square lattice ($\tau$ and $\sigma$ for Pauli matrices), in which 1D gapped topological edge states and four topological corner states are presented in the 
nontrivial phase $|\gamma / \lambda| < 1$. 
Remarkably, each plaquette in this 2D SSH model contains a $\pi$-flux.

\begin{figure}[t]
    \raggedbottom
    \begin{minipage}{0.61\linewidth}
        \raggedbottom
        \includegraphics[width=1.0\textwidth]{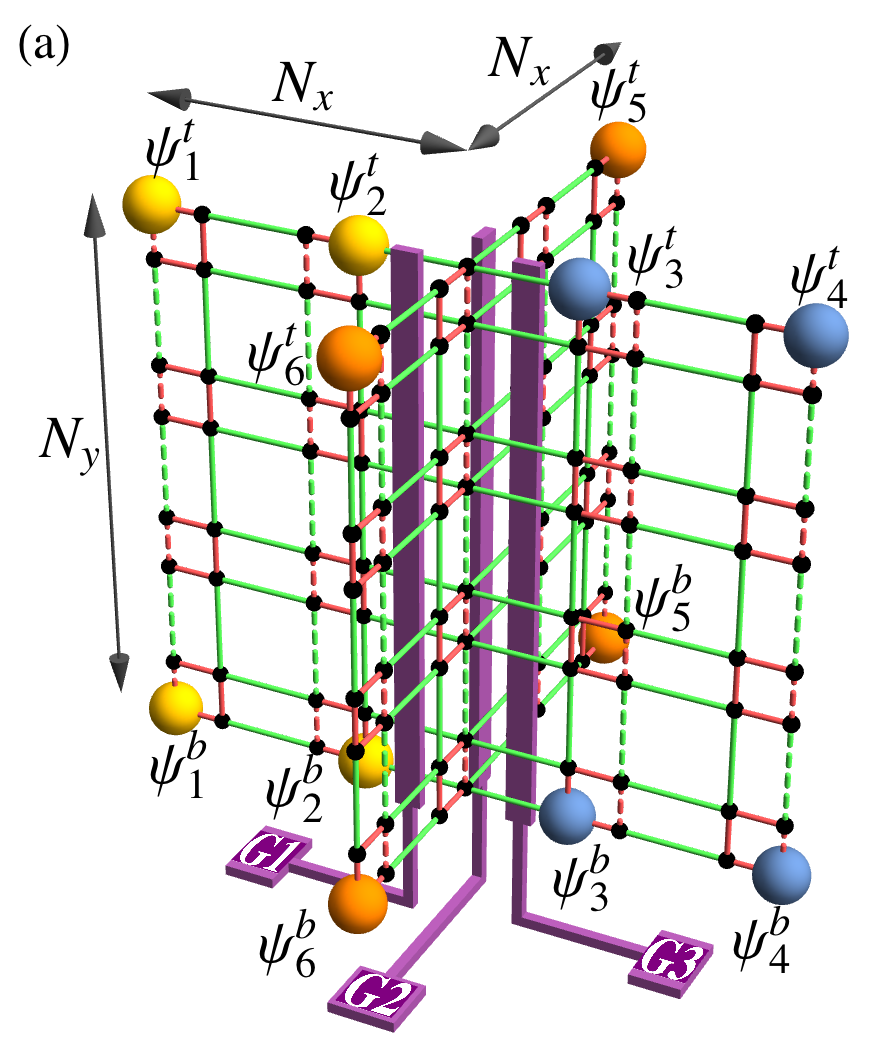}

    \end{minipage}%
    \begin{minipage}{0.38\linewidth}
        \raggedbottom
        \includegraphics[width=1.0\linewidth]{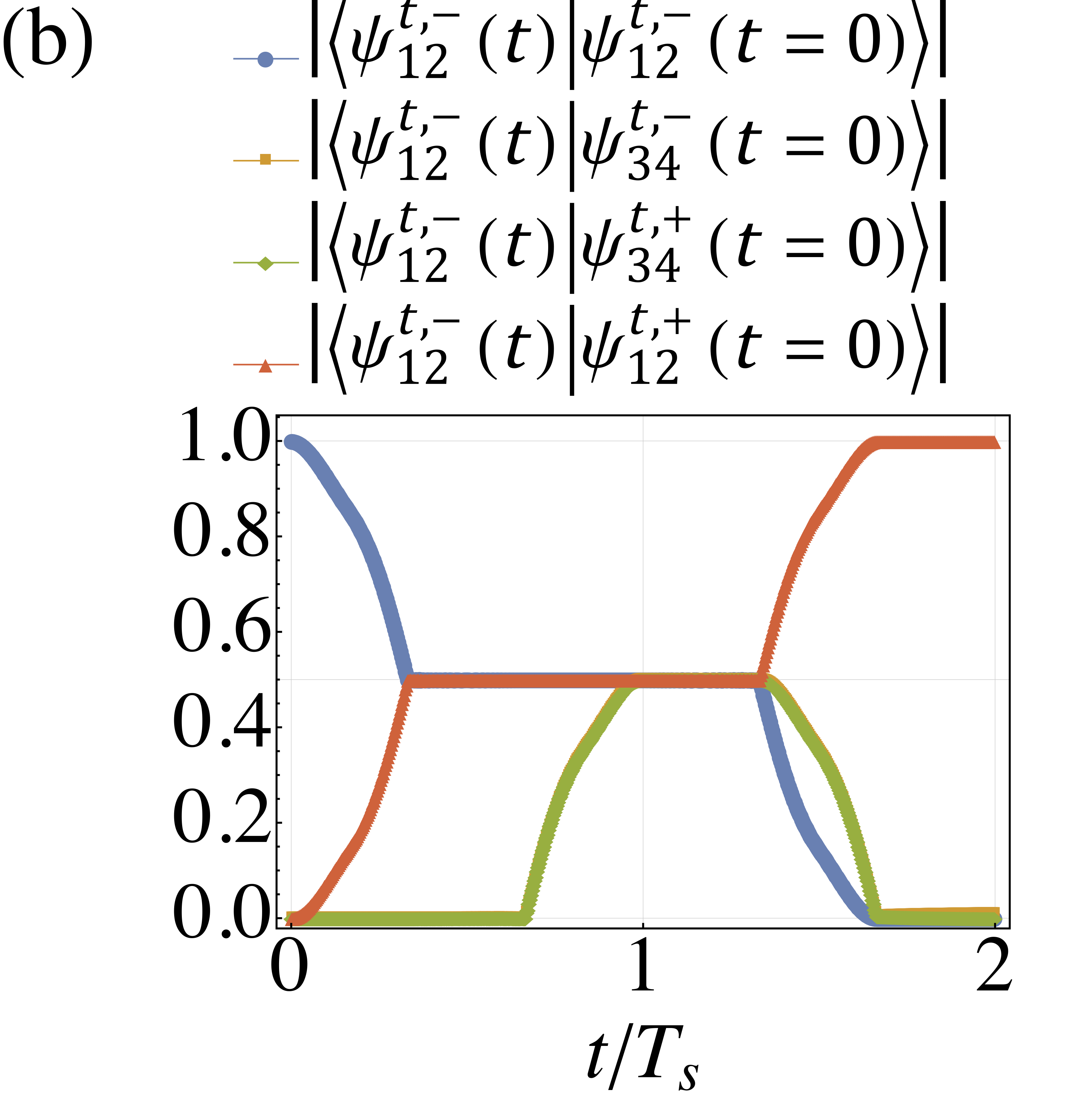}
        \includegraphics[width=1.0\linewidth]{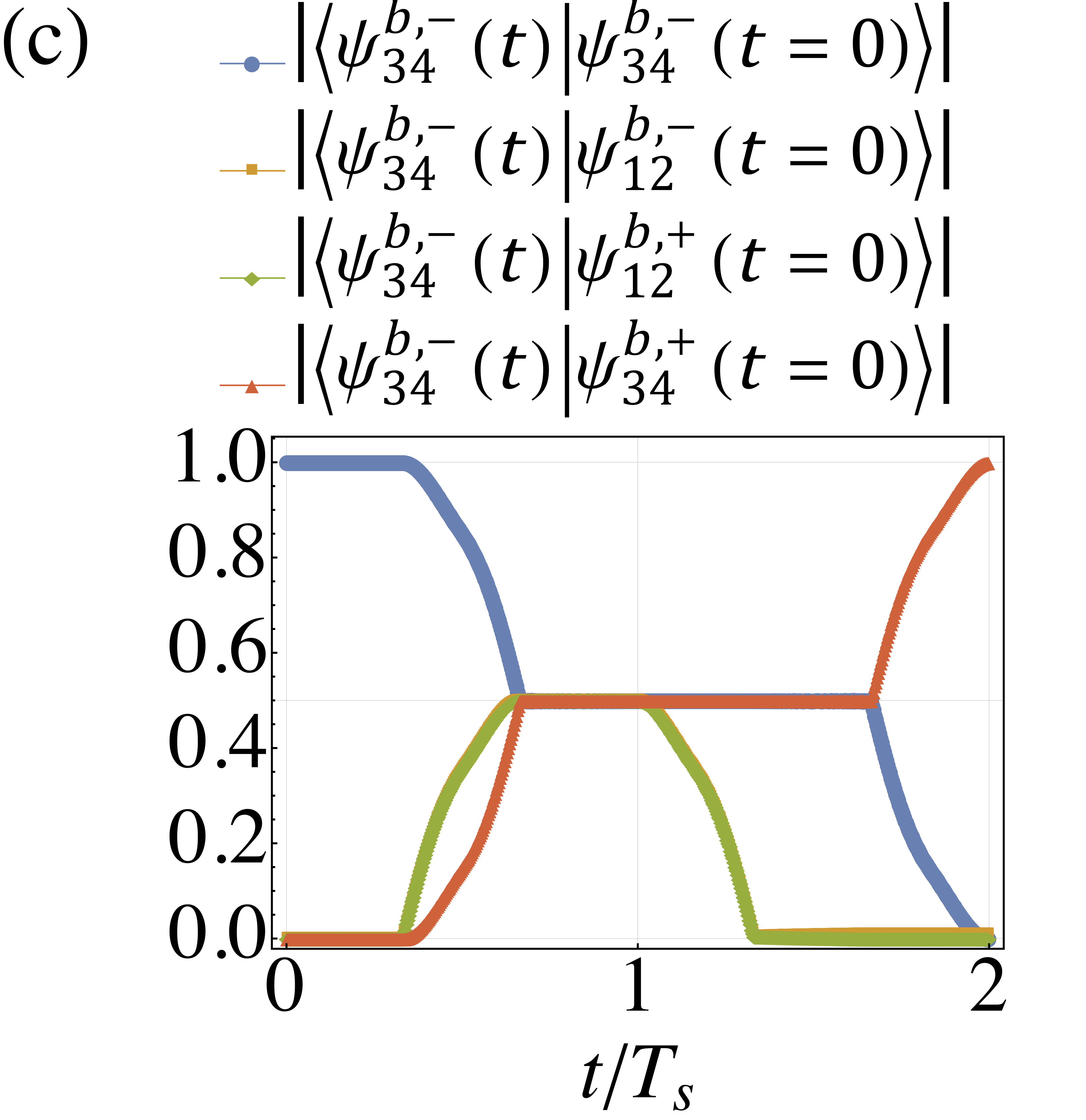}
    \end{minipage}
	
\caption{(a) Sketch of the cross-shaped junction formed by four arms each a 2D SSH square lattice in the size of $N_x \times N_y$ and three gates G1, G2, and G3. The red (green) bond denotes hopping strength $\gamma$ ($\lambda$). The dashed bond denotes hopping term with negative sign. Spatial positions for the six pairs of topological corner states before braiding are shown as large spheres (three pairs on the top edge $\psi_{i=1,2,...,6}^t$, and three pairs on the bottom edge $\psi_{i=1,2,...,6}^b$, respectively). (b), (c) Evolution of the wavefunctions (b) $| \psi_{12}^{t,-} (t) \rangle$; (c) $| \psi_{34}^{b,-} (t) \rangle$ during the whole braiding process that $\psi_{2}^{e}$ and $\psi_{3}^{e}$ ($e=t,b$) are swapped twice in succession.  Parameters of the numerical calculations in (b), (c) are $N_x=5$, $N_y=40$, $\lambda=1.0$, $\gamma=0.1$, and $T_s=1200$.}
\label{cross_junction}
\end{figure}

A cross-shaped junction supporting the braiding operation \cite{cross_junction, Chui-Zhen_cross_junction, YijiaWu} is composed of four arms each a topologically nontrivial 2D SSH lattice in the size of $N_x \times N_y$, and three voltage gates (G1, G2, and G3) located near the cross point [Fig. \ref{cross_junction}(a)]. Each arm of this junction can be isolated from others by the potential barrier induced by the presence of the corresponding gate voltage. For example, before the braiding, gate voltages in G1 and G3 are turned on, while gate voltage in G2 is turned off, therefore the cross-shaped junction is divided into three isolated parts and twelve topological corner states $\psi_{i}^{e}$ are presented as shown in Fig. \ref{cross_junction}(a) [edge index $e=t (b)$ for corner states on the top (bottom) edge. $i=1,2,...,6$]. Here we choose $N_y \gg N_x > \xi$, where $\xi$ is the localization length of the topological corner states. In this way, the coupling between the top and the bottom edge can be neglected, hence these twelve corner states are separated into two identical sets as $\psi_{i}^{t} $'s on the top edge, and $\psi_{j}^{b}$'s on the bottom edge ($i,j = 1,2,...,6$). By adiabatically \cite{SupplementaryMaterial} tuning the gate voltages in three steps \cite{cross_junction, Chui-Zhen_cross_junction, YijiaWu}, the spatial positions of two topological corner states $\psi_{2}^{t}$ and $\psi_{3}^{t}$ as well as the positions of $\psi_{2}^{b}$ and $\psi_{3}^{b}$ can be swapped simultaneously. In the first step, the gate voltage in G1 is adiabatically turned off at first, so that both $\psi_2^t$ and $\psi_2^b$ become extended states spatially distributed across two arms. After that, the gate voltage in G2 is adiabatically turned on, therefore $\psi_2^t$ ($\psi_2^b$) becomes localized on the top (bottom) edge at the back side of G2. In the second step, G3 is adiabatically turned off and then G1 is adiabatically turned on, hence $\psi_3^t$ ($\psi_3^b$) moves to the initial position of $\psi_2^t$ ($\psi_2^b$). In the final step, the swapping is accomplished by adiabatically turning off G2 then turning on G3. The time cost for such swapping process is $T_s$.

Due to the finite-size-induced coupling between $\psi_{2i-1}^{e}$ and $\psi_{2i}^{e}$ ($e=t,b$; $i=1,2,3$), the eigenstates of the cross-shaped junction before braiding ($t=0$) are symmetric or asymmetric states as $| \psi_{12}^{e,\pm} (t=0) \rangle = \frac{1}{\sqrt{2}} (| \psi_{1}^{e} \rangle \pm e^{-i\alpha_{12}^{e}} | \psi_{2}^{e} \rangle)$ and $| \psi_{34}^{e,\pm} (t=0) \rangle = \frac{1}{\sqrt{2}} (| \psi_{4}^{e} \rangle \pm e^{-i\alpha_{34}^{e}} | \psi_{3}^{e} \rangle)$, where $\alpha_{12}^{e}$ and $\alpha_{34}^{e}$ are arbitary phases \cite{DanielLoss_J-R_non-Abelian-1}. During the whole braiding process, $\psi_{2}^{e}$ and $\psi_{3}^{e}$ ($e=t,b$) are swapped twice in succession. Though all these corner states eventually come back to their initial spatial positions, the eigenstate before braiding $| \psi_{2i-1,2i}^{e,\pm} (t=0) \rangle $ evolves as $| \psi_{2i-1,2i}^{e,\pm}(t) \rangle = U(t) | \psi_{2i-1,2i}^{e,\pm}(t=0) \rangle$, where $U(t) = \hat{T} \exp[i \int_{0}^{t} \mathrm{d}\tau H(\tau)]$ is the time evolution operator ($\hat{T}$ for time-ordering operator). For example, an eigenstate $| \psi_{12}^{t,-} (t=0) \rangle$ before braiding evolves into another eigenstate as $| \psi_{12}^{t,-} (t=2 T_s) \rangle = | \psi_{12}^{t,+} (t=0) \rangle$ 
[Fig. \ref{cross_junction}(b)], implying that an additional $\pi$-phase is picked up as $\psi_{2}^{t} \to -\psi_{2}^{t}$
. In the same way, by investigating the time-evolution of other eigenstates [e.g. $| \psi_{34}^{b,-} (t) \rangle$ in Fig. \ref{cross_junction}(c)], we confirm that $\psi_{i}^{e} \to -\psi_{i}^{e}$ after $\psi_{2}^{e} $ and $\psi_{3}^{e}$ are swapped twice in succession ($e=t,b$; $i=2,3$). As a result, if $\psi_2^e$ and $ \psi_3^e $ are swapped once only, their non-Abelian nature is exhibited as $\psi_2^e \to \psi_3^e $ and $\psi_3^e \to -\psi_2^e $ (up to a gauge transformation). In brief, the topological corner states here are two identical sets of Dirac fermionic modes being braided simultaneously and exhibiting identical braiding properties.


\textit{Analytical description on the non-Abelian braiding.} 
Though the non-Abelian braiding properties of HOTI's topological corner states have been numerically demonstrated, an analytical description 
on such braiding, especially on the relation between the non-Abelian braiding and the nontrivial topology is still highly needed. We notice that the wavefuntion of the topological corner state, for example, in the lower left corner of the 2D SSH lattice and with vanishing momentum, has the form of $\psi^{\mathrm{cor}}(x,y) = C^{\mathrm{cor}} (e^{-x/\xi_+^{\mathrm{cor}}}-e^{-x/\xi_-^{\mathrm{cor}}}) (e^{-y/\xi_+^{\mathrm{cor}}}-e^{y/\xi_-^{\mathrm{cor}}}) (0,1,0,0)^T$, where $\xi_{\pm}^{\mathrm{cor}} = \frac{1\pm\sqrt{1-2(1+\gamma/\lambda)}}{2(1+\gamma/\lambda)}$ are the localization lengths ($\gamma/\lambda > -1$ is required), and $C^{\mathrm{cor}}$ is the normalization constant. The spatial part of such wavefunction 
is reminiscent of the other Dirac-type topological states such as 
the vortex-induced bound state \cite{SQShen_half_vortex, Shun-Qing_Shen_book, NuclearPhysicsB_DiracFermion} which is also a zero-dimensional localized state in 2D topological system. 
The Dirac-type bound state possessing zero-energy is presented with the half-flux vortex, which is also in parallel with the $\pi$-flux in each plaquette of the 2D SSH model. As we will show below, the 2D topological system with vortices could be served as an additional model proving the non-Abelian nature of the topologically protected Dirac fermionic modes in an analytical fashion. (In comparison, the swap of particles' spatial positions is prohibited in a strict 1D system.)

\begin{figure}[t]
    \centering
	\includegraphics[width=0.48\textwidth]{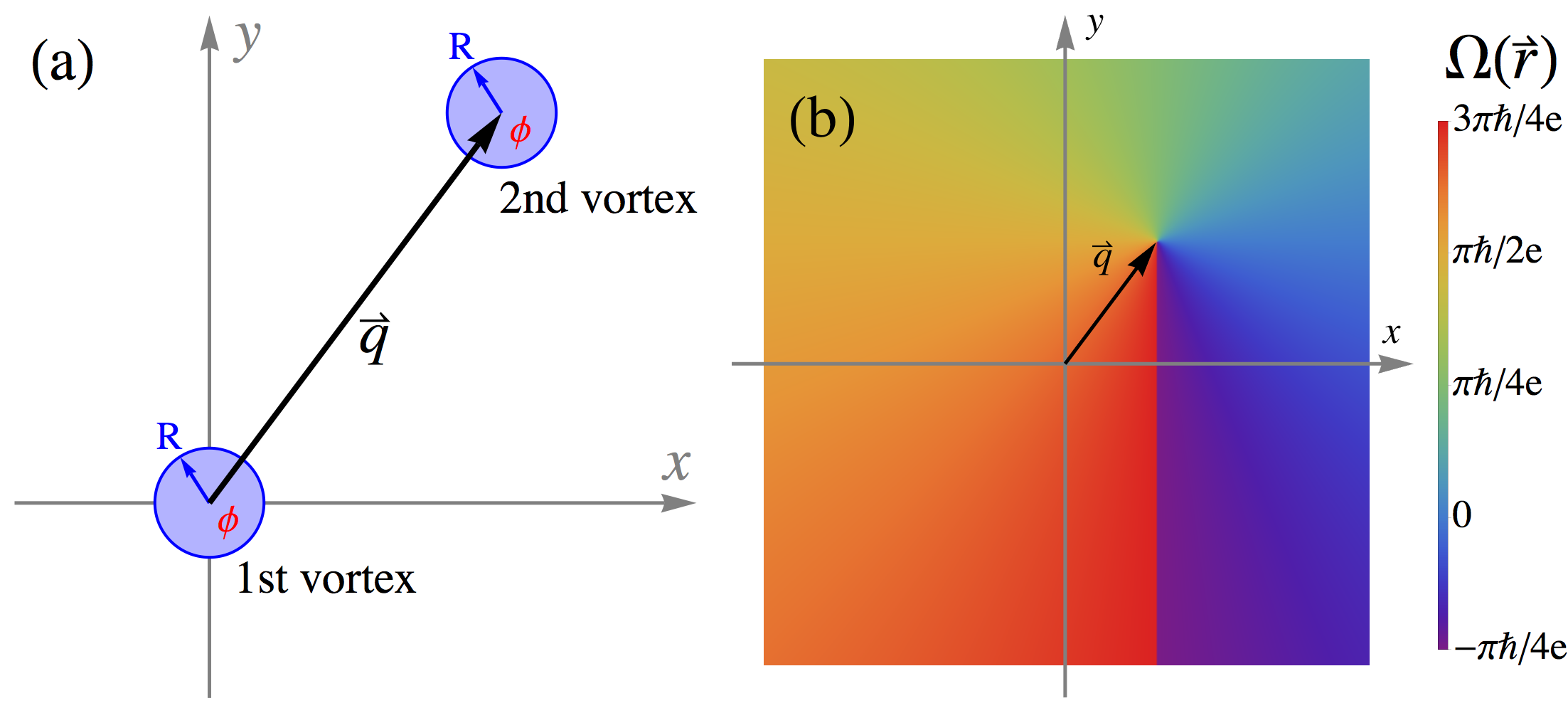}
\caption{(a) Schematic plot of two vortices with radius $R$ and magnetic flux $\phi = \phi_0/2 = \pi\hbar/e$ threaded in a QAHI. (b) Gauge field $\Omega(\mathbf{r})$ with a branch cut along the $-\hat{y}$ direction.}

\label{vortex_braiding}
\end{figure}

Specifically, considering a QAHI $h_{\mathrm{QAHI}} \left(\mathbf{p}\right) = A(p_{x}\sigma_{x}-p_{y}\sigma_{y}) + (M-B\mathbf{p}^{2})\sigma_{z}$ \cite{BHZ_model_science, BHZ_model_RMP} with two holes punched, where the first hole is placed at the origin
, while the second one is at $\mathbf{q}$ [Fig. \ref{vortex_braiding}(a)]. Both holes have a radius $R$ ($R \ll |\mathbf{q}|$) and threaded by half-flux $\phi = \phi_0 / 2$ ($\phi_0 = h/e$ for flux quantum) so that two vortices are formed. A Dirac fermionic mode \cite{Shun-Qing_Shen_book} with vanishing momentum is localized at the first vortex as $ \psi_{1}^{\mathrm{vor}}\left(\mathbf{r}\right) = \frac{C^{\mathrm{vor}}}{\sqrt{r}} [ e^{-(r-R)/\xi_+^{\mathrm{vor}}} - e^{-(r-R)/\xi_-^{\mathrm{vor}}} ] \exp[-\frac{ie}{\hbar}\Omega\left(\mathbf{r}\right)] \cdot (e^{i\theta}, i)^T $, where $\xi_{\pm}^{\mathrm{vor}} = \frac{A\pm\sqrt{A^{2}-4MB}}{2M}$ are the localization lengths and $C^{\mathrm{vor}}$ is the normalization constant \cite{SupplementaryMaterial}.
The vector potential induced by the second vortex has been taken into consideration as the gauge field \cite{SupplementaryMaterial}

\begin{equation}
\Omega\left(\mathbf{r}\right) = \begin{cases}
\frac{\hbar}{2e}\cdot\arctan\left[\frac{\left(\mathbf{r}-\mathbf{q}\right)\cdot\hat{y}}{\left(\mathbf{r}-\mathbf{q}\right)\cdot\hat{x}}\right] & , \left(\mathbf{r}-\mathbf{q}\right)\cdot\hat{x}>0\\
\frac{\hbar}{2e}\cdot\left\{ \arctan\left[\frac{\left(\mathbf{r}-\mathbf{q}\right)\cdot\hat{y}}{\left(\mathbf{r}-\mathbf{q}\right)\cdot\hat{x}}\right]+\pi\right\}  & , \left(\mathbf{r}-\mathbf{q}\right)\cdot\hat{x}<0
\end{cases}
\label{gauge_field}
\end{equation}

\noindent where $\hat{x}$, $\hat{y}$ denotes the unit vector along the $x$ and $y$ direction, respectively. There is a branch cut \cite{IvanovPRL2001} of the gauge field $\Omega\left(\mathbf{r}\right)$ along the $-\hat{y}$ direction [Fig. \ref{vortex_braiding}(b)] and the phase jump across the branch cut is $\pi\hbar/e$.

Similarly, the wavefunction of the Dirac fermionic mode bounded with the second vortex is denoted as $\psi_{2}^{\mathrm{vor}} \left(\mathbf{r}-\mathbf{q}\right)$, in which the gauge field induced by the first vortex has also been included. If the (relative) spatial positions of these two vortices are swapped through a counterclockwise rotation, then the first Dirac fermionic mode passes through the branch cut of the second vortex \cite{IvanovPRL2001} and acquires an additional $\pi$ phase, while the second Dirac fermionic mode does not go through the branch cut of the first vortex. Therefore, we analytically obtain the braiding properties of these Dirac fermionic modes as

\begin{equation}
\begin{cases}
\psi_{1}^{\mathrm{vor}} \left(\mathbf{r}\right) \to -\psi_{2}^{\mathrm{vor}}\left(\mathbf{r}-\mathbf{q}\right)\\
\psi_{2}^{\mathrm{vor}} \left(\mathbf{r}-\mathbf{q}\right) \to \psi_{1}^{\mathrm{vor}}\left(\mathbf{r}\right)
\end{cases},
\label{swapping_rule}
\end{equation}

\noindent which is exactly the same as the topolgocial corner states in HOTI as expected. This analytical derivation unambiguously demonstrates that the non-Abelian braiding here comes from the flux-induced geometric phase. We conclude that non-Abelian braiding can also be exhibited for Dirac fermionic modes, provided that nontrivial topology \cite{vonOppen_braiding, Pachos_review} is presented. Remarkably, identical non-Abelian behaviors are also presented for half-flux vortices in quantum spin Hall insulator (2D TI) \cite{SQShen_half_vortex} which is two copies of QAHIs related by time-reversal (TR) symmetry.


\textit{Braiding operator for Dirac fermionic mode.} Based on the investigation on both the topological corner states in HOTI and the Dirac-type bound states in QAHI, we have shown that an operation $T_{i}$ swapping two Dirac fermionic modes $\psi_{i}$ and $\psi_{i+1}$ will give rise to $\begin{cases}
\psi_{i}\to\psi_{i+1}\\
\psi_{i+1}\to-\psi_{i}\\
\psi_{j}\to\psi_{j}
\end{cases}$ ($j\neq i$ and $j\neq i+1$), which is reminiscent of the braiding properties of MZM \cite{IvanovPRL2001} and bosonic mode \cite{nonAbelian_light}. Nevertheless, the braiding operator $\tau\left(T_{i}\right)$ for Dirac fermionic modes obeying $\tau\left(T_{i}\right) \psi_{j} \left[\tau\left(T_{i}\right)\right]^{-1} = T_{i}\left(\psi_{j}\right)$
has the explicit form of \cite{SupplementaryMaterial}

\begin{equation}
\tau\left(T_{i}\right) = \exp \left[ \frac{\pi}{2} \left( \psi_{i+1}^{\dagger}\psi_{i}  - \psi_{i}^{\dagger}\psi_{i+1} \right) \right].
\label{braiding_operator}
\end{equation}

\noindent In comparison, the braiding operators are $\tau (T_i) = \exp [ \frac{\pi}{2} (b_{i+1}^{\dagger}b_{i} - b_{i}^{\dagger}b_{i+1}) ] $ for bosonic mode \cite{nonAbelian_light} and $\tau(T_i) = \exp [ \frac{\pi}{4} \gamma_{i+1} \gamma_i ]$ for MZM \cite{IvanovPRL2001}, where  $b_i$ and $\gamma_i$ are bosonic and Majorana operators, respectively. If each Dirac fermionic mode is decomposed into two MZMs with different ``flavors'' as $\psi_i \equiv \frac{1}{2} (\gamma_i^a + i \gamma_i^b)$ ($a$, $b$ for flavor indices), then the equivalent form of Eq. (\ref{braiding_operator}) as $\tau(T_i) = \exp [ \frac{\pi}{4} \gamma_{i+1}^a \gamma_i^a ] \exp [ \frac{\pi}{4} \gamma_{i+1}^b \gamma_i^b ]$ is the tensor product of Majorana braiding operator with different flavors.


Such tensor product form implies that the crucial difference between the non-Abelian statistics of the MZM and the Dirac fermionic mode lies in their quantum dimensions. For instance, for a Majorana system composed of four MZMs, the quantum dimension is $2^2=4$ and its Hilbert space could be divided into two two-dimensional sectors with different fermion parities. All the braiding operators  are block diagonal for the braiding operations conserving the fermion parity \cite{IvanovPRL2001}. In contrast, the complete Hilbert space for a fermionic system composed of four Dirac fermionic modes ($\psi_1$, $\psi_2$, $\psi_3$, and $\psi_4$) is in the quantum dimension of $2^4=16$ \cite{NuclearPhysicsB_DiracFermion}. Such a Hilbert space can also be divided into five sectors each labeled by a different fermion number from zero to four. The braiding operators $\tau(T_i)$ conserving fermion number have different forms in each of these sectors. Considering that these four fermionic modes are coupled in the manner of $\psi_{12}^{\pm} = \frac{1}{\sqrt{2}} (\psi_1 \pm e^{-i\alpha_{12}}\psi_2)$ and $\psi_{34}^{\pm} = \frac{1}{\sqrt{2}} (\psi_4 \pm e^{-i\alpha_{34}}\psi_3)$, then the four-dimensional single-fermion sector can be spanned as $\left[ (\psi_{12}^{-})^{\dagger} |0\rangle, (\psi_{12}^{+})^{\dagger} |0\rangle, (\psi_{34}^{-})^{\dagger} |0\rangle, (\psi_{34}^{+})^{\dagger} |0\rangle \right] ^{T}$. In such a basis, the braiding operators 
are presented in the matrix form as $\tau\left(T_{1}\right) = \left(\begin{array}{cc}
i\sigma_z & 0\\
0 & \sigma_0
\end{array}\right)$, $\tau\left(T_{3}\right) = \left(\begin{array}{cc}
\sigma_0 & 0\\
0 & i\sigma_z
\end{array}\right)$ ($\sigma$ for Pauli matrix), and

\begin{equation}
\tau\left(T_{2}\right)=\frac{1}{2}\left(\begin{array}{cccc}
1 & 1 & 1 & -1\\
1 & 1 & -1 & 1\\
-1 & 1 & 1 & 1\\
1 & -1 & 1 & 1
\end{array}\right),
\label{tau_T2}
\end{equation}

\noindent respectively (without loss of generality, here we set $\alpha_{12} = \alpha_{34} = \pi/2$) \cite{SupplementaryMaterial}. All of the two-qubit Pauli rotations \cite{Pachos_book_TQC} $\sigma_i \otimes \sigma_j$ ($i,j = 0,1,2,3$) can be implemented (up to an overall phase) by combining swapping operations $T_1$, $T_2$, and $T_3$. Such two-qubit Pauli rotations are naturally the combination of the single-qubit operations on each set of MZMs with different flavors. It is worth noting that the $\tau(T_2)$ in Eq. (\ref{tau_T2}) is consistent to the numerical results on the braiding of HOTI's topological corner states [Fig. \ref{cross_junction}(b), (c)] when $\psi_2^e$ and $\psi_3^e$ ($e=t,b$) are swapped once ($t=T_s$). Finally, $\tau\left(T_{2}\right)$ neither commutes with $\tau\left(T_{1}\right)$ nor $\tau\left(T_{3}\right)$, indicating that they cannot be diagonalized simultaneously.
Such non-diagonal braiding matrix mathematically
displays the non-Abelian nature of the Dirac fermionic modes in an unambiguous way.


\textit{Realizing non-Abelian braiding in topological circuit.} Experimentally, it is generally difficult to manipulate the spatial positions of vortices in 2D topological systems. In comparison, the braiding operations are relatively easier through tuning gate voltages in quasi-2D structures such as trijunction or cross-junction \cite{MatthewFisher_T-junction, cross_junction, JayDSau_trijunction}. Moreover, the HOTI supporting topological corner states has been realized in topological circuits \cite{2D_SSH_circuit_1, 2D_SSH_circuit_2, 3D_SSH_circuit_experiment}. Here, we propose an alternative circuit scheme, in which not only the 2D SSH model, but also the non-Abelian braiding of the topological corner states could be implemented.

\begin{figure}[t]
    \centering
	\includegraphics[width=0.45\textwidth]{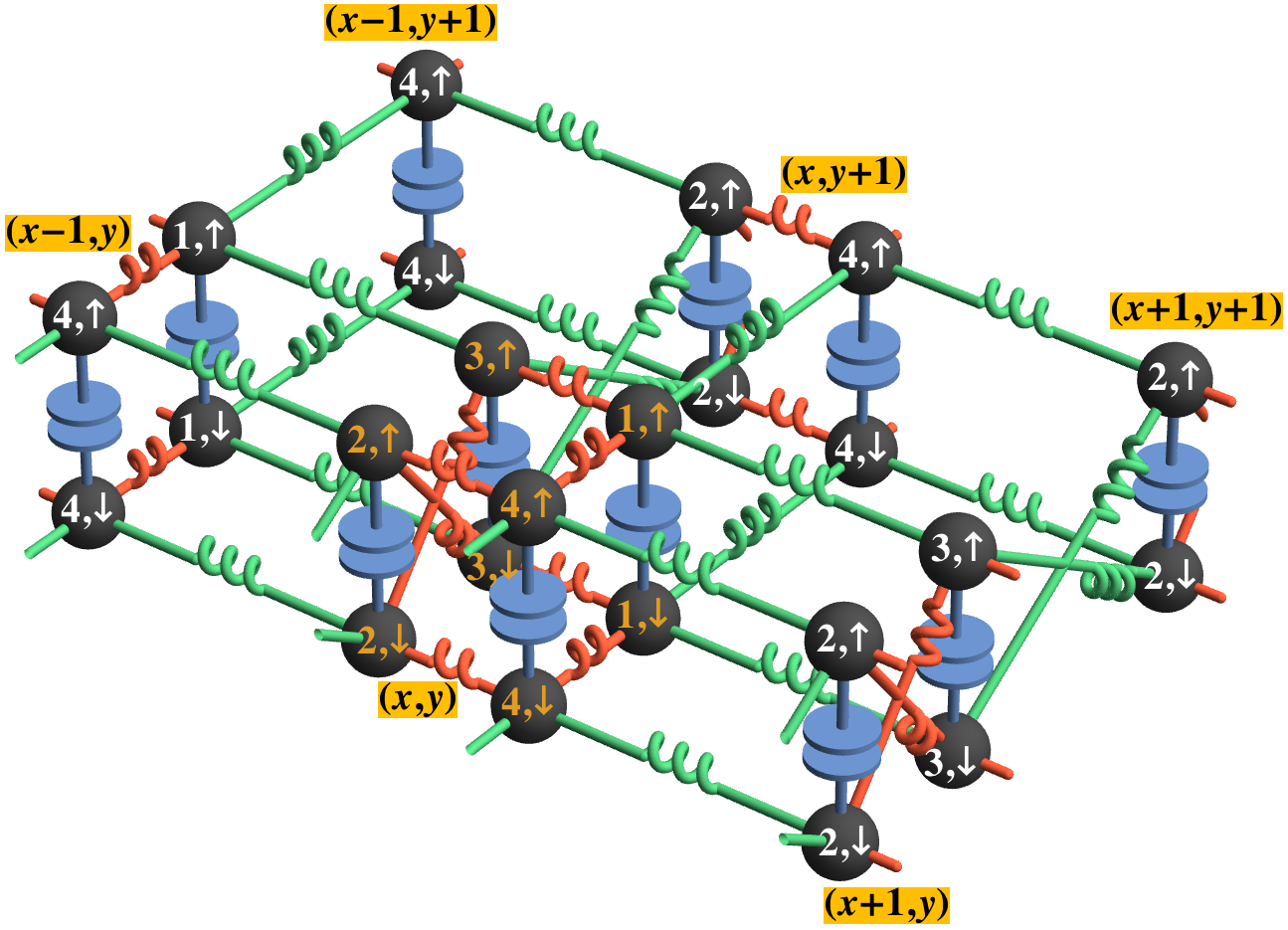}
\caption{Sketch of the topological circuit network realizing 2D SSH model. Each unit cell is labeled with spatial coordinates $(x,y)$. The nodes with orange text denote one of the unit cells. The red (green) screw indicates inductor with inductance $L_{\gamma}$ ($L_{\lambda}$). The parallel plates in blue indicates capacitor with capacitance $C$. The grounding inductor $L_x$ for unit cells in each row is not shown.}
\label{LC_circuit}
\end{figure}

As shown in Fig. \ref{LC_circuit}, the circuit network is in a bilayer structure and each unit cell labeled by spatial coordinates $(x,y)$ contains eight nodes denoted by $(n,s)$, where the sublattice index $n = 1,2,3,4$ and the layer index $s = \uparrow,\downarrow$. The two nodes with opposite layer indices in the same lattice sites are connected through a capacitor with capacitance $C$, while the nearst neighbouring lattice sites are alternately connected by inductor with inductance $L_{\gamma}$ or $L_{\lambda}$. Moreover, the cross-connection of inductor is periodically adopted to generate the $\pi$-flux \cite{2D_SSH_circuit_2, circuit_flux_1, circuit_flux_2, ZhiqiangZhang} in the 2D SSH model. Finally, all nodes inside the unit cells in the same row (same coordinate $x$) is grounded with an inductor with inductance $L_{x}$.

The current flowing into each node $I_{x,y}^{n,s} (\omega)$ is related to the voltage in each node $V_{x,y}^{n,s} (\omega)$ by Kirchhoff's law as $I_{x,y}^{n,s} (\omega) = \sum J_{x,x'; y,y'}^{n,n'; s's'} (\omega) V_{x',y'}^{n',s'} (\omega)$, in which $J(\omega)$ is the circuit Laplacian \cite{Zhao_topological_circuits, RuiYu_nodal_line_LC} and $\omega$ is the frequency of the alternating current (AC). Through a unitary transformation $\left(\begin{array}{c}
V_{x,y}^{n}\\
U_{x,y}^{n}
\end{array}\right)=\frac{1}{\sqrt{2}}\left(\begin{array}{cc}
1 & -1\\
1 & 1
\end{array}\right)\left(\begin{array}{c}
V_{x,y}^{n,\uparrow}\\
V_{x,y}^{n,\downarrow}
\end{array}\right)$, one find that $U_{x,y}^{n}$ does not contain oscillation term and therefore can be dropped \cite{circuit_flux_1, ZhiqiangZhang}. Denoting eigenenergy $E \equiv 2\omega^{2}C$, on-site energy $\epsilon_{x} \equiv \frac{2}{L_{\lambda}}+\frac{2}{L_{\gamma}}+\frac{1}{L_{x}}$, hopping strength $\gamma \equiv \frac{1}{L_{\gamma}}$ and $\lambda \equiv \frac{1}{L_{\lambda}}$, then the explicit expression for $V_{x,y}^{n}$

\begin{eqnarray}
E V_{x,y}^{1} &=& \epsilon_{x} V_{x,y}^{1}-\lambda\left(V_{x+1,y}^{3}+V_{x,y+1}^{4}\right)-\gamma\left(V_{x,y}^{3}+V_{x,y}^{4}\right) \nonumber \\
E V_{x,y}^{2} &=& \epsilon_{x} V_{x,y}^{2}-\lambda\left(V_{x-1,y}^{4}-V_{x,y-1}^{3}\right)-\gamma\left(V_{x,y}^{4}-V_{x,y}^{3}\right) \nonumber \\
E V_{x,y}^{3} &=& \epsilon_{x} V_{x,y}^{3}-\lambda\left(V_{x-1,y}^{1}-V_{x,y+1}^{2}\right)-\gamma\left(V_{x,y}^{1}-V_{x,y}^{2}\right) \nonumber \\
E V_{x,y}^{4} &=& \epsilon_{x} V_{x,y}^{4}-\lambda\left(V_{x+1,y}^{2}+V_{x,y-1}^{1}\right)-\gamma\left(V_{x,y}^{1}+V_{x,y}^{2}\right) \nonumber \\
& &
\end{eqnarray}

\noindent is in parallel with the lattice Hamiltonian describing the 2D SSH model \cite{SupplementaryMaterial}, where $V_{x,y}^{n}$ plays the role of the wavefunction at the corresponding lattice site. Experimentally, for example, if we choose $C = 100\mathrm{pF}$, $L_x = 1\mathrm{mH}$, $L_{\lambda} = 1\mathrm{mH}$, and $L_{\gamma} = 10 L_{\lambda}$ ($\gamma = 0.1 \lambda$), then the resonant frequency for the topological corner states is $\omega = 4\mathrm{MHz}$.

The cross-shaped junction in Fig. \ref{cross_junction} could be constructed based on such topological circuit network. The three gates (G1, G2, and G3 in Fig. \ref{cross_junction}) inducing potential barriers are implemented by replacing the corresponding grounding inductor $L_x$ near the cross point of the junction by variable grounding inductors. By tuning down the inductance of the grounding variable inductors, the on-site energy $\epsilon_{x}$ for all the lattice sites in the corresponding row increases simultaneously in the manner of $\epsilon_{x} \equiv \frac{2}{L_{\lambda}}+\frac{2}{L_{\gamma}}+\frac{1}{L_{x}}$, hence a potential barrier is formed. In this way, the braiding operation swapping topological corner states $\psi_{2}^{e}$ and $\psi_{3}^{e}$ ($e = t, b$) could be performed in topological circuit through tuning three variable inductors in succession \cite{cross_junction, Chui-Zhen_cross_junction, YijiaWu}.



\textit{Conclusions.} We have demonstrated the non-Abelian braiding properties of topological corner state, a kind of Dirac fermionic mode protected by nontrivial topology in HOTI. Such braiding operation may be realized through cross-shaped junction constructed by topological electric circuit. The non-Abelian nature of the topologically protected Dirac fermionic mode is proved to be highly related to its nontrivial topology. The braiding operator as well as the braiding matrix have also been explicitly expressed for the Dirac fermionic mode.


\textit{Acknowledgements.} We thank Chui-Zhen Chen, Jin-Hua Gao, Qing-Feng Sun, and Zhi-Qiang Zhang for fruitful discussion. This work is financially supported by the National Basic Research Program of China (Grants No. 2015CB921102, No. 2017YFA0303301, and No. 2019YFA0308403) and the National Natural Science Foundation of China (Grants No. 11534001, No. 11674028, No. 11822407, and No. 11974271).

\bibliography{2D_HOTI}

\end{document}